\documentclass[letterpaper]{article} % DO NOT CHANGE THIS
\usepackage{aaai25}  % DO NOT CHANGE THIS
\usepackage{times}  % DO NOT CHANGE THIS
\usepackage{helvet}  % DO NOT CHANGE THIS
\usepackage{courier}  % DO NOT CHANGE THIS
\usepackage[hyphens]{url}  % DO NOT CHANGE THIS
\usepackage{graphicx} % DO NOT CHANGE THIS
\usepackage{natbib}  % DO NOT CHANGE THIS AND DO NOT ADD ANY OPTIONS TO IT
\usepackage{caption} % DO NOT CHANGE THIS AND DO NOT ADD ANY OPTIONS TO IT
\usepackage{amsmath}
\usepackage{amsfonts}
\usepackage{algorithm}
\usepackage{algorithmic}

\usepackage{newfloat}
\usepackage{listings}
\DeclareCaptionStyle{ruled}{labelfont=normalfont,labelsep=colon,strut=off} % DO NOT CHANGE THIS
\floatstyle{ruled}
\newfloat{listing}{tb}{lst}{}
\floatname{listing}{Listing}
\title{Bridging Training and Execution via Dynamic Directed Graph-Based Communication in Cooperative Multi-Agent Systems}
\author{
    Zhuohui Zhang\textsuperscript{\rm 1,\rm 2},
    Bin He\textsuperscript{\rm 1,\rm 2},
    Bin Cheng\textsuperscript{\rm 1,\rm 2}\thanks{Corresponding author},
    Gang Li\textsuperscript{\rm 1,\rm 2}
}
\affiliations{
    \textsuperscript{\rm 1}Department of Control Science \& Engineering, Tongji University,  China\\
    \textsuperscript{\rm 2}National Key
Laboratory of Autonomous Intelligent Unmanned Systems, Tongji University,  China\\
    \{zhangzhuohui, hebin, bincheng, lig\}@tongji.edu.cn

}

\usepackage{bibentry}
\begin{document}

\maketitle

\begin{abstract}

Multi-agent systems must learn to communicate and understand interactions between agents to achieve cooperative goals in partially observed tasks. However, existing approaches lack a dynamic directed communication mechanism and rely on global states, thus diminishing the role of communication in centralized training. Thus, we propose the Transformer-based graph coarsening network (TGCNet), a novel multi-agent reinforcement learning (MARL) algorithm. TGCNet learns the topological structure of a dynamic directed graph to represent the communication policy and integrates graph coarsening networks to approximate the representation of global state during training. It also utilizes the Transformer decoder for feature extraction during execution. Experiments on multiple cooperative MARL benchmarks demonstrate state-of-the-art performance compared to popular MARL algorithms. Further ablation studies validate the effectiveness of our dynamic directed graph communication mechanism and graph coarsening networks.

\end{abstract}

\begin{links}
    \link{Code}{https://github.com/ZhuohuiZhang/TGCNet}
\end{links}

\section{Introduction}
\label{S_1}

Cooperative multi-agent reinforcement learning (MARL) problems have emerged in the past decade as a vital framework for addressing intricate collaborative challenges in real-world scenarios \cite{wei2019colight, vinyals2019grandmaster, cheng2024discrete, li2023distributed}, attracting considerable attention and showcasing substantial potential for practical applications and commercial viability. A natural approach to cooperative MARL regards the multi-agent system as a whole using centralized methods \cite{tan1993multi}, which face scalability challenges and limitations associated with centralized controllers \cite{foerster2018counterfactual}. Another approach is decentralized, where each agent learns its policy with single agent techniques \cite{tampuu2017multiagent}, addressing the limitations of centralized controllers but introducing non-stationarity and credit assignment problems \cite{lowe2017multi}. To mitigate these issues, decentralized policies can be learned using the paradigm of centralized training with decentralized execution (CTDE) \cite{sunehag2017value, yuan2022multi, yu2022surprising, zhang2019efficient}, which includes value-based \cite{rashid2020monotonic, eccles2019biases, wang2021towards} and policy-based \cite{yu2022surprising, lin2021learning} frameworks. CTDE-based MARL algorithms rely on access to the global state, which is often an idealized assumption.

Communication is essential in multi-agent systems for sharing information, learning, and collaborating toward common goals, especially in partially observable environments. It is crucial for complex tasks such as coordinating autonomous vehicles \cite{cao2012overview}, sensor networks \cite{pipattanasomporn2009multi}, and multi-robot control \cite{fox2000probabilistic}. Effective communication underpins cooperation. Recently, inspired by human cooperation, communication has been integrated into MARL to enhance information sharing among agents. Early work \cite{foerster2016learning, sukhbaatar2016learning} disseminated information through a broadcast format, but this approach led to high communication costs and information redundancy. Subsequent research has aimed to reduce communication overhead by selectively determining when to communicate \cite{singh2018learning} and eliminating redundant information through targeted peer-to-peer communication \cite{jiang2018learning, jiang2018graph}. However, there is a lack of a universal communication method that can address the following five issues simultaneously with (1) whom to communicate, (2) when to communicate, (3) which piece of information to communicate, (4) how to combine and integrate the information received, and (5) how to use communication to avoid reliance on global state are limited.

In this work, we focus on developing communication mechanisms in MARL to enhance team performance on collaborative tasks, while minimizing communication costs and avoiding reliance on global state information. We propose a novel MARL algorithm, the Transformer-based graph coarsening network (TGCNet), which models cooperative agents' communication as dynamic directed graphs, with each agent as a node. The adjacency trajectory matrix represents the structure of the dynamic directed graph. We connect the training and execution phases through dynamic directed graphs, allowing agents to communicate during both stages. TGCNet integrates a Transformer-based \cite{vaswani2017attention} multi-key gated communication mechanism with the Q-network, enabling end-to-end training without the need for additional loss functions. The multi-key gated mechanism learns through a hard additive attention approach \cite{bahdanau2014neural}, using multi-keys for aggregation to determine the number of communications between agents. Based on the dynamic directed graph, we introduce a graph coarsening network that utilizes self-attention pooling and coarsening operations to approximate the global state. We conduct a comprehensive empirical study across diverse cooperative multi-agent benchmarks to evaluate the efficacy of our proposed methodology. Specifically, we evaluate our method on the Hallway scenario \cite{wang2019learning}, the Level-Based Foraging (LBF) environment \cite{rangwala2020learning}, and eight distinct maps from the StarCraft Multi-Agent Challenge (SMAC) \cite{samvelyan2019starcraft}. In addition, we perform component analyses to demonstrate the effectiveness of the multi-key gated communication network and the graph coarsening network. The contributions of this study are as follows: 

\begin{itemize}

\item We formalize a novel general paradigm for communicative and cooperative MARL, which bridges training and execution through dynamic directed graphs. The communication policy is represented by the structure of dynamic directed graphs, which is shared during both training and execution.

\item We design a multi-key gated communication network, which learns the structure of a dynamic directed graph. It can achieve multiple peer-to-peer directed communications between agents within the same time step, an efficient mechanism for information transmission.

\item During centralized training, graph coarsening networks process the mix network, while in distributed execution, a Transformer architecture handles the Q-network; both phases leverage learned communication policies as inputs for global state coarsening aggregation and information communication with feature extraction, respectively.

\end{itemize}

\section{Related Works}

Communication is vital for MARL to capture the dependencies between agent actions. In addition, it has been demonstrated to effectively improve exploration and team rewards \cite{apicella2012social}. Our work builds on and relates to previous research on MARL and communication mechanisms. The existing MARL can be divided into two categories based on whether or not a communication mechanism is set up. MARL without communication uses the CTDE paradigm, which has been successfully implemented with value-based and policy-based algorithms. The focus of policy-based MARL is to stabilize training with centralized state value estimation. Representative works in this area include COMA \cite{foerster2018counterfactual}, MADDPG \cite{lowe2017multi}, and MAPPO \cite{yu2022surprising}. The other category is value-based MARL, following the Individual-Global-Max (IGM) principle; it focuses on value function factorization. This approach includes VDN \cite{sunehag2017value}, QMIX \cite{rashid2020monotonic}, QTRAN \cite{son2019qtran}, and QPlex \cite{wang2020qplex}. Although these algorithms have shown significant performance in many multi-agent cooperative tasks, their effectiveness relies heavily on the introduction of global state and the setup of centralized trainer. Unlike existing works, our algorithm facilitates communication among agents, thereby eliminating reliance on global state.

The second category focuses on the use of communication mechanisms to compensate for the limitations of local observations. Previous works enforce fixed and static broadcasting for communication. RIAL and DIAL \cite{foerster2016learning} are applied to improve communication skills by broadcasting messages across time steps. CommNet \cite{sukhbaatar2016learning} broadcasts hidden states as messages and obtains fused information by averaging. IC3Net \cite{singh2018learning} includes a gating mechanism to learn when to broadcast using CommNet. BiCNet \cite{peng2017multiagent} and ATOC \cite{jiang2018learning} consider bidirectional communication and use bidirectional RNNs \cite{zaremba2014recurrent} or LSTMs \cite{greff2016lstm} to achieve bidirectional broadcast. Recent studies aim to develop more sophisticated communication mechanisms to prevent indiscriminate message broadcasting. DGN \cite{jiang2018graph} leverages graph convolutional networks with relational kernels to capture dynamic agent interactions. TarMAC \cite{das2019tarmac} adopts an attention mechanism to determine whether two agents must communicate and distinguishes the importance of incoming messages. G2ANet \cite{liu2020multi} constructs a communication interaction graph using a two-stage attention mechanism. However, these algorithms only facilitate communication before agents take actions, neglecting communication during the centralized training process. We propose a new multi-key gated communication network for bidirectional, modeling communication as a dynamic directed graph, combined with Transformer for information fusion and retrieval, to address these concerns. Moreover, we integrate graph neural networks to convert the communication flow into a structured network of dynamic directed graphs that change dynamically at each time step, and leverage graph coarsening network to approximate the global state.

\section{Preliminaries}
\label{s_3}

\subsubsection{Decentralized Partially Observable Markov Decision Processes (Dec-POMDP)}

We model cooperative multi-agent systems as Dec-POMDP \cite{oliehoek2016concise}, which imposes partially observable settings on multi-agent Markov decision processes. A Dec-POMDP can be described by a tuple $\langle \mathcal{N}, \mathcal{S}, \mathcal{A}, P, \Omega, O, R, \gamma, \mathcal{M} \rangle$, where $\mathcal N = \{1, \cdots ,n\}$ indicates the set of $n$ agents, $\mathcal{S}$ denotes the set of global states, $\mathcal{A}$ refers to the set of actions, $P$ represents the state transition function, $\Omega$ denotes the set of observations, $O$ refers to the observation function, $R$ represents the reward function, $\gamma \in [0, 1)$ indicates the discount factor, and $\mathcal{M}$ is the set of messages that agents can communicate. At each time step $t$, each agent $i \in \mathcal{N}$ receives an observation $o_t^i \in \Omega$ from the observation function $O(s_t, i)$, where $s_t \in \mathcal{S}$. Each agent follows an individual policy $\pi^i(a_t^i | \tau_t^i, m_t^i)$, where $\tau_t^i = (o_1^i, a_1^i, \ldots, o_{t-1}^i, a_{t-1}^i, o_t^i)$ is the action-observation history of agent $i$, and $m_t^i \in \mathcal{M}$ is the message received by agent $i$ at time $t$. The joint action $\boldsymbol{a}_t = \langle a_t^1, \ldots, a_t^n \rangle$ leads to the next state $s_{t+1} \sim P(s_{t+1} | s_t, \boldsymbol{a}_t)$, and the team receives a global reward $R(s_t, \boldsymbol{a}_t)$. The objective is to find a joint policy $\boldsymbol{\pi} = \langle \pi^1, \ldots, \pi^n \rangle$ that maximizes the global action-value function $Q_{\mathrm{tot}}^{\boldsymbol{\pi}}(\boldsymbol{\tau},\boldsymbol{a}) = \mathbb{E}_{s,\boldsymbol{a}}\left[\sum_{t=0}^{\infty}\gamma^tR(s,\boldsymbol{a})\mid s_0=s,\boldsymbol{a}_0=\boldsymbol{a},\boldsymbol{\pi}\right]$, with $\boldsymbol{\tau}=\langle\tau_1,\ldots,\tau_n\rangle$. Each agent serves as both sender and receiver.

\subsubsection{Dec-POMDP with Dynamic Directed Graph} 
\begin{figure}
  \centering
  \includegraphics[width=0.8\columnwidth]{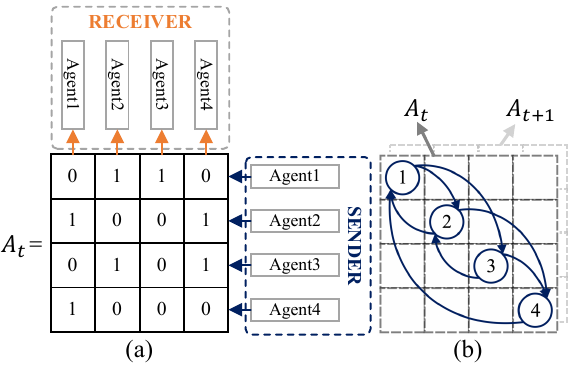}
  \caption{Dynamic directed graph. (a) Adjacency trajectory matrix at time $t$. (b) Dynamic directed graph. The dynamic directed graph can represent the associations and communication structures between nodes at a certain moment.}
  \label{f_1}
\end{figure}
Based on Dec-POMDP, we propose a communication topology using dynamic directed graphs. The structural properties of the graph are represented by the adjacency trajectory matrix $\mathbf{A}$, which is denoted as $\mathbf{A}\in \mathbb{B}^{l\times n\times n}$, where $\mathbb{B}$ indicates the Boolean matrix, $l$ denotes the length of the trajectory and $n$ indicates the number of agents. If $\lceil A_{t}^{ij} \rceil = 0$, then no communication occurred from agent $i$ to agent $j$ at time $t$. If $\lceil A_{t}^{ij} \rceil = 1$, then communication transpired from agent $i$ to agent $j$ at time $t$. As an example, the adjacency trajectory matrix $A_t$ at time $t$ is shown in Figure \ref{f_1}(a), and the corresponding dynamic directed graph is shown in Figure \ref{f_1}(b). After traversing the dynamic directed graph to agent $i$, the received message is computed as $\tilde{m}_{t}^{i} = A_{t}^{i} \odot m_t^{i}$, where $\odot$ means element-wise multiplication. For value-based MARL, which uses the action-value function to update policies, where in the distributed execution phase, each agent learns a Q-Network $Q^i(\tau,a,\tilde{m}; \theta)$ \cite{mnih2015human} to approximate the action-value function $Q^i(s, a)$. In the centralized training phase, the group of agents learns a mix network $Q_{\mathrm{tot}}(\boldsymbol{\tau},\boldsymbol{a},\boldsymbol{\tilde{m}},s;\boldsymbol{\theta})$ to approximate the global action value function $Q_{\mathrm{tot}}(s,\boldsymbol{a})$. The parameters $\boldsymbol{\theta}$ are learned by minimizing the expected temporal difference (TD) error:

\begin{equation}
\label{e_1}
    \mathcal{L}(\boldsymbol{\theta})=\sum_{i=1}^b\left[\left(y_i^{\mathrm{tot}}-Q_{\mathrm{tot}}(\boldsymbol{\tau}_t,\boldsymbol{a}_t,\boldsymbol{\tilde{m}}_t,s_t;\boldsymbol{\theta})\right)^2\right],
\end{equation} where $b$ is the batch size of transitions sampled from the replay buffer $\mathcal{D}$  and $y^{\mathrm{tot}}=R(s_t, \boldsymbol{a}_t) +\gamma\max_{\boldsymbol{a}_{t+1}}Q_{\mathrm{tot}}(\boldsymbol{\tau}_{t+1},\boldsymbol{a}_{t+1},\boldsymbol{\tilde{m}}_{t+1},s_{t+1};\boldsymbol{\theta}^{-})$, $\boldsymbol{\theta}^{-}$ are the parameters of a target network.

\section{Method}

In this section, we introduce the detailed structure and design of TGCNet. Fundamentally, TGCNet is an extension of value-based MARL. Our original intention is for TGCNet to simultaneously solve the five communication problems for MARL, specifically in Section \ref{S_1}. Our motivation stems from the objective of communication in MARL, which is to mitigate the negative effects of limited observations in the Dec-POMDP. Through communication, agents can obtain an approximation of the global state. Our aim is to learn a state function $S(o^i_t, m^i_t, A^i_t)$. Moreover, if an agent's information significantly affects the global state for agent $i$ during centralized training, it should also be crucial during decentralized execution. Communication plays two roles in TGCNet. In the Q-network, the agent uses communication to acquire information beyond its immediate observations to assist in decision making, and the message $m_t^i$ at time $t$ defined as the action-observation history of all other agents except agent $i$, denoted as $\tau^{-i}_t$. In the mix network, agents coarsen and aggregate individual local observations into global states through communication and the message $m_t^i$ at time $t$ defined as the local observations of all other agents except agent $i$, denoted as $o^{-i}_t$. In addition, the network is scalable to scenarios where agents may change and can be plugged into any CTDE structure.

\begin{figure*}
  \centering
  \includegraphics[width=0.8\textwidth]{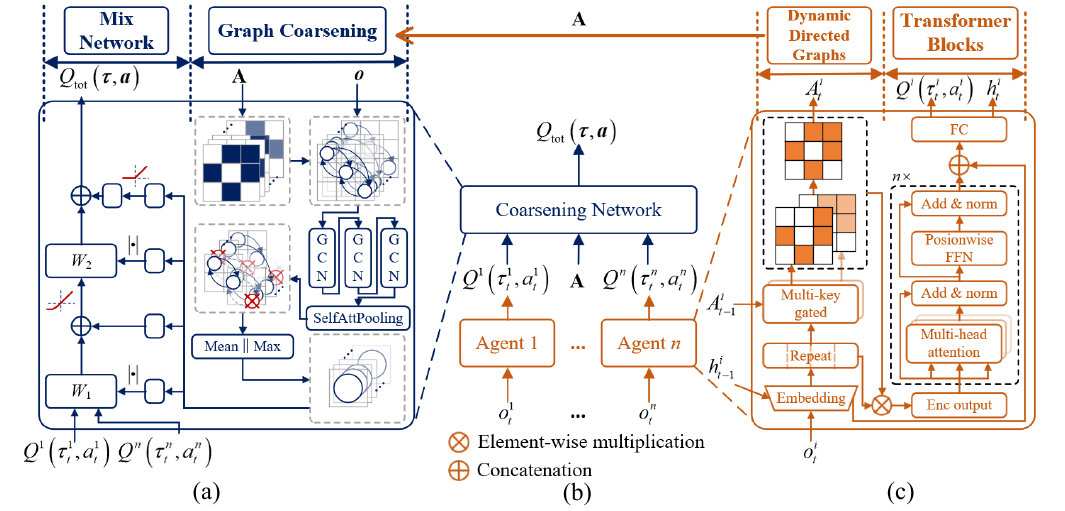}
  \caption{The network structure diagram of TGCNet. (a) Graph coarsening network and mix network. The inputs for the graph coarsening network include the local observations of the agents and the adjacency trajectory matrix. (b) Overall TGCNet architecture. (c) Transformer-Based multi-key gated communication mechanism. The communication mechanism outputs not only the individual state value function $Q^i(\tau^i,a^i)$ and hidden variables $h_t^i$ but also the adjacency trajectory matrix $A^i_{t}$. Then, this matrix is passed into the graph coarsening network to complete end-to-end backward propagation updates.}
  \label{f_2}
\end{figure*}

\subsection{Theoretical Analysis}
\label{S_4_1}
We first theoretically analyze how the state function is constructed in the mix network. We construct a 3D dynamic directed which is defined as $\mathbf{G}=(T, V, E)$, where $t \in T$ is the set of the trajectory in the time dimension, $v^{i}\in V$ denotes the set of vertex, which is composed of agents and $e^{i,j}\in E$ denotes the set of edges connecting agent $i$ and agent $j$. A dynamic directed graph can be regarded as a composition of $T$ subgraphs $G_t=(V_t, E_t)$ organized along a temporal trajectory. We define the adjacency trajectory matrix to describe the connections (communication) associations between graph vertices. The adjacency trajectory matrices for the dynamic directed graph and subgraphs are denoted by $\mathbf{A}$ and $A_t$. When a node can access the observations of other nodes, it can approximate the global state. However, this information may be either independent or partially redundant. We define the relationship between the global state and local observation using the following equation:

\begin{equation}
\label{eq_2}
\begin{aligned}
s_{t} = \text{Agg}\left(o_t^{j},\forall\nu^{j}\in N(\nu^{i}) \parallel o_t^i\right) \\ - \text{Overlap}\left(o_t^{j},\forall\nu^{j}\in N(\nu^{i}) \parallel o_t^i\right),
\end{aligned}
\end{equation}
where $\text{Agg}\left(\cdot \right)$ indicates an aggregator function, $N(\nu^{i})$ represents the set of neighbor nodes of node $i$, $\text{Overlap}\left(\cdot \right)$ denotes an overlap function and $\parallel$ represents the operator of parallel concatenation, which connects two vectors in parallel. It is evident that $o_t^{j},\forall\nu^{j}\in N(\nu^{i}) \parallel o_t^i$ is essentially equivalent to $\tilde{m}^i_t$. When dealing with aggregation and overlap functions, it is crucial to maintain consistent output results regardless of the order of neighboring nodes. We design the summation aggregation functions $\text{Agg}^\text{sum}$ and pooling overlap functions $\text{Overlap}^\text{pool}$, as follows:

\begin{equation}
\label{eq_3}
    \mathrm{Agg}^{\mathrm{sum}}=\sigma\left(\mathrm{sum}\ \left\{ W\left[\tilde{m}^i_t \parallel o_t^i\right]+b \right\} \right),
\end{equation}
\begin{equation}
\label{eq_4}
    \mathrm{Overlap}^\mathrm{pool}=\mathrm{max}\left\{\sigma\left(W\left[\tilde{m}^i_t \parallel o_t^i\right]+b\right)\right\},
\end{equation}
where $\sigma$ denotes the activation function, $W$ and $b$ represent the weights and bias of a feed-forward neural network.

After constructing the state function, we only need to replace $s$ from Section \ref{s_3} to obtain the global action value function of TGCNet, as definded $Q_{\mathrm{tot}}(\boldsymbol{\tau},\boldsymbol{a},\boldsymbol{\tilde{m}},\text{Agg}(\boldsymbol{\tilde{m}}\parallel \boldsymbol{o}) - \text{Overlep}(\boldsymbol{\tilde{m}}\parallel \boldsymbol{o});\boldsymbol{\theta})$. The expression of $Q_{\mathrm{tot}}$ shows that it excludes the global state as an input. The network design schematics are shown in Figure \ref{f_2}.

\begin{figure*}[t]
  \centering
  \includegraphics[width=0.75\textwidth]{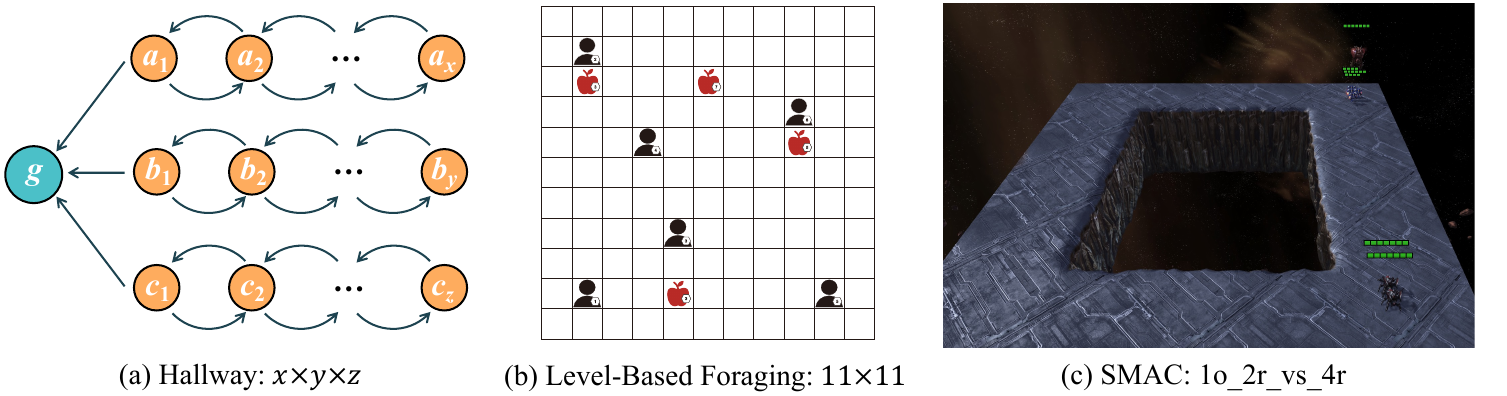}
  \caption{Multiple benchmarks used in our experiments.}
  \label{f_3}
\end{figure*}
\subsection{Graph Coarsening Network}

We construct the graph coarsening network based on Equation (\ref{eq_2}) to address the last of the five communication problems outlined in Section \ref{S_1}. The schematics of the graph coarsening network are shown in Figure \ref{f_2}(a). The dynamic directed graph, composed of multiple subgraphs, can be considered a coarsened graph, where each subgraph is a super node to fit the global state. This approach enables hierarchical learning of global information. According to the theoretical analysis in Section \ref{S_4_1}, the input of the graph coarsening network consists of each agent's local observations $o^i_t$, the adjacency trajectory matrix $A^i_t$ and the received message $\tilde{m}^i_t$. We normalize the adjacency trajectory matrix to prevent degree bias and gradient vanishing during training. We define $[\tilde{m}^i_t \parallel o^i_t]$ as $X^i_t$ as feature inputs for each node in graph convolutional networks (GCN). The specific network structure is expressed as follows:

\begin{equation}
    \tilde{X^i_t} = \sigma \left(\tilde{D^i_t}^{-1/2} \tilde{A^i_t} \tilde{D^i_t}^{-1/2} X^i_t \right),
\end{equation} where $\tilde{A^i_t}$ represents the adjacency matrix with self-loops $\tilde{A^i_t} = A^i_t + I$, $I$ means identity matrix and $\tilde{D^i_t}$ represents its degree matrix. Following three iterations of graph convolution, the output is fed into self attention pooling (SelfAttPooling) \cite{lee2019self}. The structure is as follows:

\begin{equation}
    {\tilde{X^i_t}^{\prime}} = \tilde{X^i_t} \odot \tanh \left( \tilde{D^i_t}^{-1/2} \tilde{A^i_t} \tilde{D^i_t}^{-1/2} \tilde{X^i_t}\right).
\end{equation}

Based on Equations (\ref{eq_3}) and (\ref{eq_4}), we have developed a readout mechanism that performs a one-time aggregation operation on all nodes, resulting in a globally coarsened representation of the graph that approximates the global state function. This readout mechanism is similar to the global pooling operation commonly used after the last convolutional layer in CNN models \cite{lecun1998gradient}. Both approaches aggregate all inputs into a global representation in a single step, expressed as follows:

\begin{equation}
    s_t=\sigma\left(W\left[\frac{1}{N}\sum_{i=1}^{n}\tilde{X^i_t}^{\prime}\parallel\max_{i=1}^{n}\tilde{X^i_t}^{\prime}\right]+b\right).
\end{equation}

After approximating the global state using graph coarsening networks, we implement the output of $Q_{\mathrm{tot}}$ in Section \ref{S_4_1} using a similar approach to QMIX \cite{rashid2020monotonic}.

\begin{figure*}[t]
  \centering
  \includegraphics[width=\textwidth]{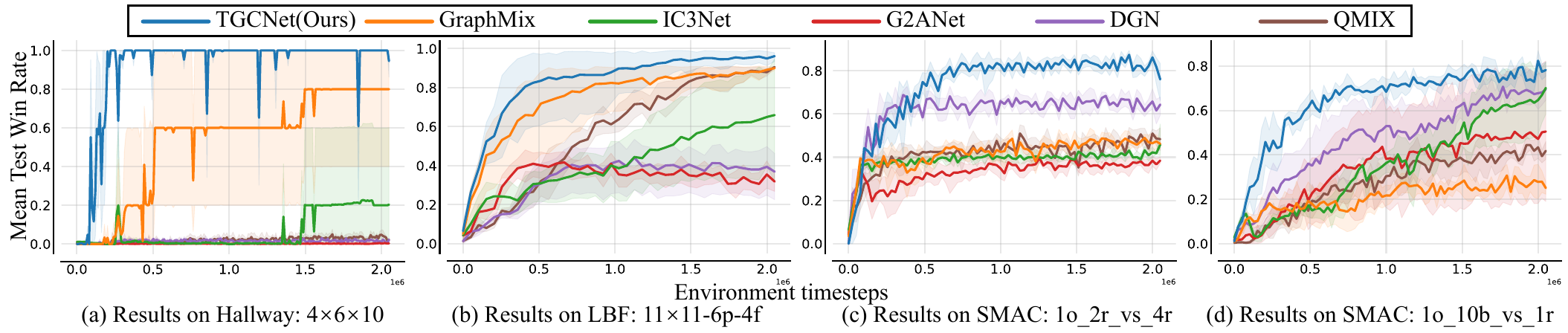}
  \caption{Performance comparison with baselines on multiple benchmarks.}
  \label{f_4}
\end{figure*}

\subsection{Transformer-Based Multi-Key Gated Communication Mechanism}
\label{S_4_2}

After constructing $S(o^i_t, m^i_t, A^i_t)$, we design a multi-key gated mechanism to learn $A^i_t$ in the Q-network. Combined with the Transformer decoder, it addresses the first four of the five communication problems discussed in Section \ref{S_1}. The multi-key gated communication network is the encoder, and the Transformer is the decoder. The schematics of the communication mechanism are shown in Figure \ref{f_2}(c).

Historical information is initialized, and the observations of each agent are pre-processed by embedding their local observations $o_t^i$ and historical information $h_{t-1}^{i}$. Then, the adjacency trajectory matrix $A^i_{t-1}$ is initialized and all its elements are set to 1, except the agent $i$ itself, which is masked off. We combine the initialized adjacency trajectory matrix and the information of each agent to perform a pre-communication, which indicates that each agent repeats the hidden variables of the other agents. The pre-communication message $\tilde{m}^i_{t-1} = A^i_{t-1} \odot m^i_{t}$ is regarded as input and processed by the multi-key gated communication network. The frequency of communication among the agents can be determined by the number of keys in the multi-key gated communication network. The output of each key $k^{i}_t$ represents the updated adjacency trajectory matrix in this communication. The specific calculation equation is as follows:

\begin{equation}
\begin{aligned}
k^{i}_t=\text{gumbel-softmax}\left(W_v^\top\tanh(W_q\tilde{m}^i_{t-1} + W_k\tilde{m}^i_{t-1})\right),
\end{aligned}
\end{equation} 
where $\text{gumbel-softmax}$ \cite{jang2016categorical} is an activation function for two output dimensions (communicate or not). After passing through communication network, the adjacency trajectory matrix is expressed as follows:

\begin{equation}
A^i_{t}= \lceil\mathop{\max}\limits_{j}k^i_t\rvert_{j})\rceil,
\end{equation}
where $k^i_t\rvert_{j}$ represents the total of $j$ keys obtained by agent $i$ at time $t$ and $\lceil \cdot \rceil$ represents rounding up. The transmission and reception of message are completed by using the updated adjacency trajectory matrix, combined with pre-communication information. The output of the encoder $\tilde{m}^i_{t} = A^i_{t} \odot m^i_{t}$ is used as input for the Transformer decoder. Each Transformer block is composed of self-attention mechanism, position-wise feed-forward networks, residual connection and layer normalization. Two Transformer blocks are commonly used to achieve an improved balance between performance and memory consumption. The output of the Transformer blocks and the previous history information concatenated together go through a fully connected layer to obtain the individual state value function $Q^i$ and the hidden state at the next moment $h_t^i$.

\begin{figure*}[t]
  \centering
  \includegraphics[width=0.95\textwidth]{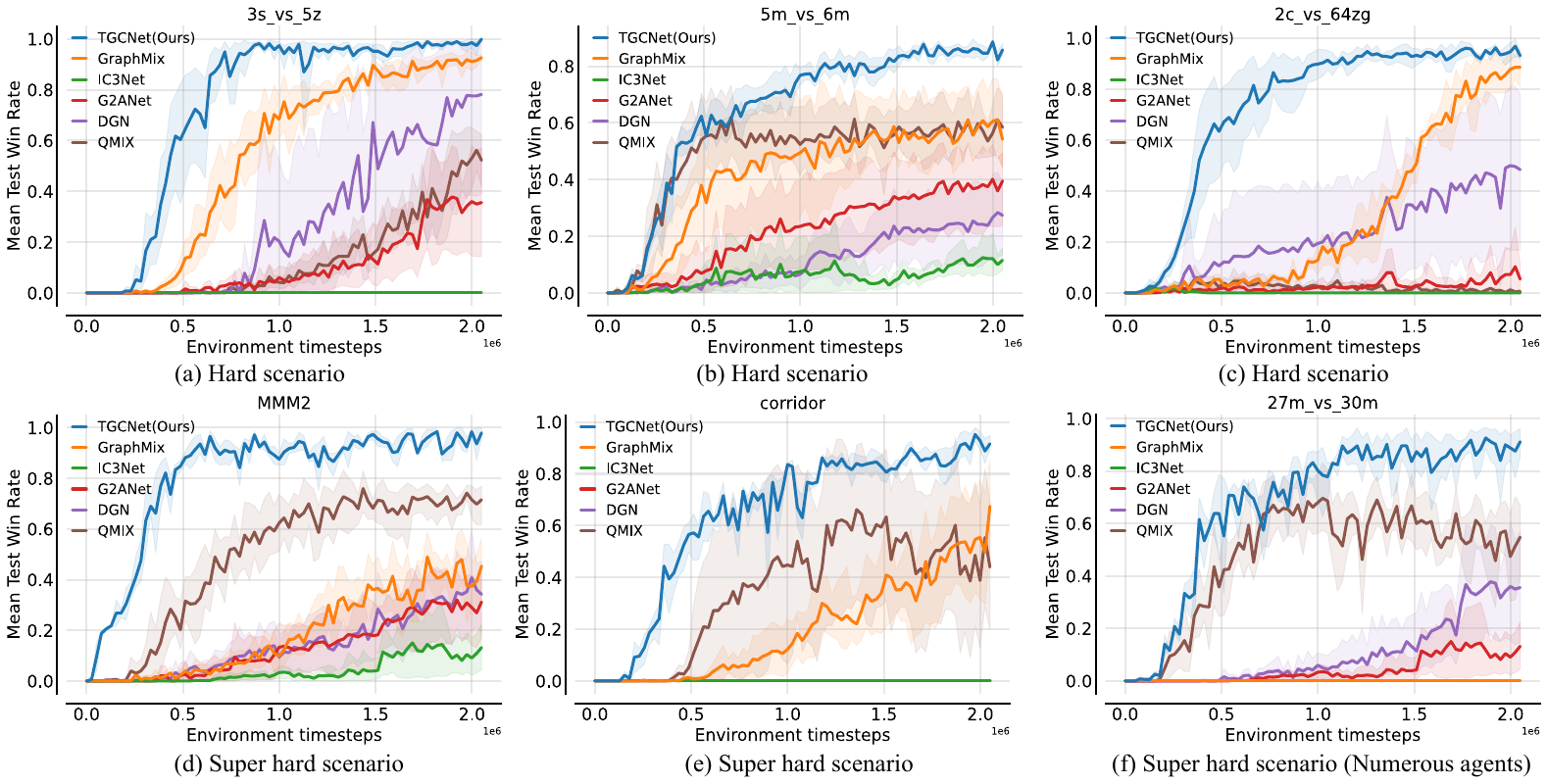}
  \caption{Performance comparison with baselines on hard (first row) and super hard (second row) scenarios in SMAC.}
  \label{f_5}
\end{figure*}

\begin{figure*}[t]
  \centering
  \includegraphics[width=0.95\textwidth]{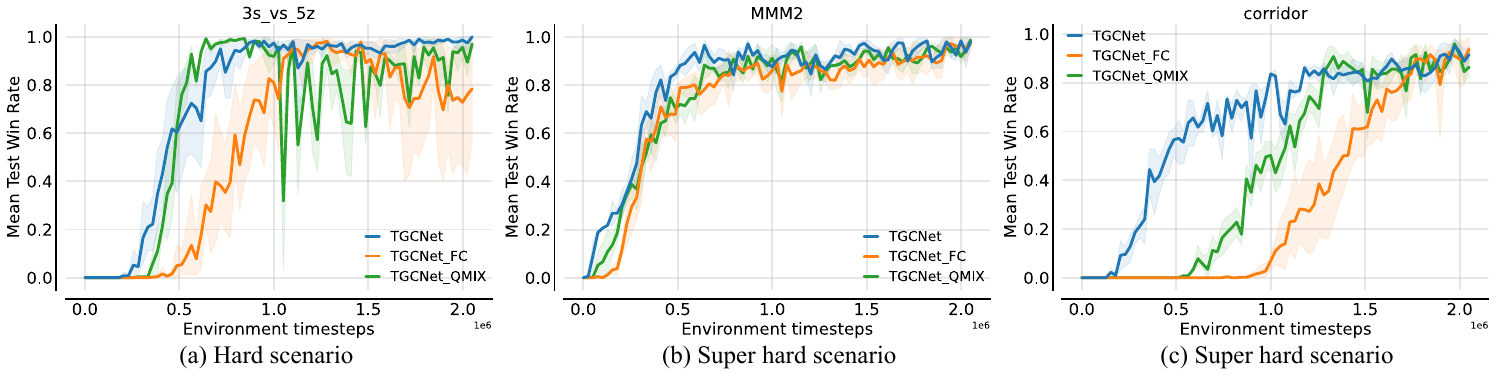}
  \caption{Ablation study results of TGCNet, TGCNet\_FC, and TGCNet\_QMIX on three hard and super hard scenarios.}
  \label{f_6}
\end{figure*}

\section{Experiments}

In this section, we evaluate our approach against five representative MARL baselines on various benchmarks with differing communication requirements. QMIX \cite{rashid2020monotonic} is a strong non-communication baseline, showing outstanding performance across multiple multi-agent benchmarks\cite{papoudakis2020benchmarking}. IC3Net \cite{singh2018learning} includes a gating mechanism for learning when to broadcast messages, using CommNet \cite{sukhbaatar2016learning}. DGN \cite{jiang2018graph} uses graph convolutional networks with relational kernels to capture dynamic agent interactions. G2ANet \cite{liu2020multi} builds a sparse communication interaction graph with a two-stage attention mechanism. GraphMix \cite{naderializadeh2020graph} also constructs a communication model using graph structures and is a notably effective baseline. All algorithms use the EPyMARL \cite{papoudakis2020benchmarking} implementation. We evaluate TGCNet with multiple state-of-the-art baselines on tasks including Hallway \cite{wang2019learning}, LBF \cite{papoudakis2020benchmarking}, and the SMAC \cite{samvelyan2019starcraft}. Figure \ref{f_3}(a) shows the Hallway task, where four agents start at different states ($a_1$ to $a_x$, $b_1$ to $b_y$, and $c_1$ to $c_z$, with $x, y, z = 4, 6, 10$) and must reach the goal $g$. Figure \ref{f_3}(b) depicts the LBF task, where six agents cooperate to collect four portions of food in an $11\times11$ grid world. In SMAC, the 1o2r\_vs\_4r and 1o10b\_vs\_1r maps \cite{wang2019learning} require collaboration and communication among agents to identify enemy positions, as shown in Figure \ref{f_3}(c). For fair evaluation, all experiments are conducted with five random seeds, and results are presented as means with a 95\% confidence interval.

\begin{figure*}[t]
  \centering
  \includegraphics[width=0.73\textwidth]{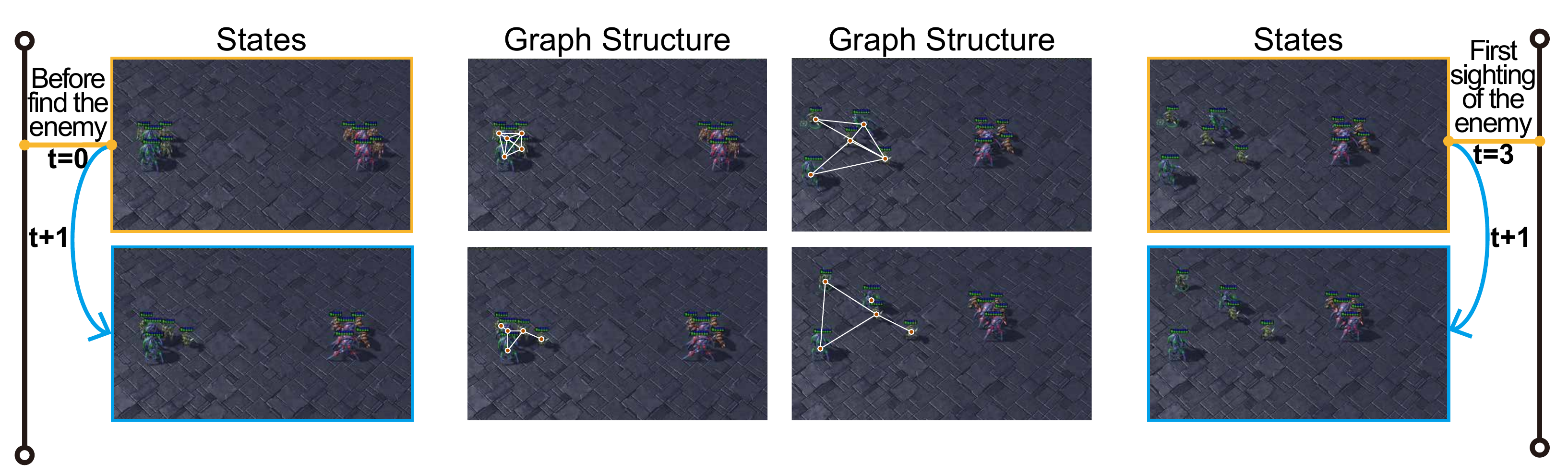}
  \caption{Information Completion visualization and analysis along the MARL trajectory.}
  \label{f_7}
\end{figure*}

\subsection{Overall Performance Comparison}

We start by evaluating the overall performance of TGCNet against multiple baselines in communication-intensive benchmarks. As shown in Figure \ref{f_4}, TGCNet consistently outperforms the other methods with low variance across all benchmarks, demonstrating its robustness in scenarios of varying difficulty. In the Hallway task (Figure \ref{f_4}(a)), which requires frequent communication for success, methods lacking communication capabilities, such as QMIX, fail. Other communication-based methods also perform poorly or fail entirely in this environment. This finding indicates that inappropriate message generation or selection would injure the learning process. The performance of TGCNet exceeds GraphMix by approximately 20\% given its effective communication modeling and powerful feature extraction capabilities. In the LBF task (Figure \ref{f_4}(b)), existing communication-based MARL methods, such as IC3Net, G2ANet, and DGN encounter challenges due to sparse rewards, especially when the food items are widely distributed. Different from its performance in Hallway, QMIX performs effectively in LBF because agents can observe nearby grids. Our method matches the performance of GraphMix, demonstrating strong coordination abilities even in scenarios with sparse rewards. For scenarios requiring communication in SMAC, QMIX performs the worst without a communication mechanism, whereas TGCNet can maintain high efficiency of learning and exhibit consistently competitive performance when converged, superior to other baselines.

To validate our algorithm's information extraction and scalability, we test it in SMAC scenarios without communication, including hard and super hard levels. As shown in Figure \ref{f_5}, TGCNet achieves the highest average test win rate across all scenarios upon convergence. GraphMix performs slightly worse than TGCNet but has a faster convergence speed and higher win rate than DGN and G2ANet due to its graph-based communication information aggregation. IC3Net, DGN, and G2ANet perform worse than QMIX in scenarios without communication because their inaccurate communication modeling fails to manage redundant information effectively, impacting the training of reinforcement learning policy networks. In hard scenarios, TGCNet's average test win rate is nearly 20\% higher than those of algorithms. This finding is attributed to the Transformer-based decoder's superior representation of policy networks. In super hard scenarios with more agents, especially on the 27m\_vs\_30m map, TGCNet maintains a higher convergence speed due to its multi-key gated communication network. Overall, TGCNet demonstrates excellent performance in all scenarios, learning effective communication strategies through its Transformer-based multi-key gated communication mechanism and accurately adjusting the global state with the graph coarsening network.

\subsection{Ablation Studies}

To understand the superior performance of TGCNet, we conduct ablation studies to evaluate the contributions of its two main components, addressing the following questions: (1) Is the graph coarsening network effective for fitting the global state? How does it compare to using real global states? (2) How does the algorithm’s effectiveness differ between full and partial communication? To address the first question, we create the TGCNet\_QMix algorithm, replacing TGCNet's graph coarsening network with a hybrid network using QMIX, while retaining other network structures. For the second question, we create TGCNet\_FC, which removes TGCNet's multi-key communication network and uses full communication (FC). We conduct ablation experiments on one hard scenario and two super hard scenarios. The results shown in Figure \ref{f_6} illustrate that TGCNet performs similarly to TGCNet\_QMIX in these scenarios, indicating that TGCNet's graph coarsening network effectively approximates the global state. TGCNet\_FC converges more gradually than TGCNet in the corridor scenario but ultimately achieves the same performance level. This finding suggests that the multi-key communication network optimizes the communication structure without sacrificing performance, thereby avoiding communication waste.

\subsection{Communication Performance}

To analyze the communication strategies learned through the graph structure, we conduct a visualization analysis of the multi-agent communication network after training, focusing on a test scenario in SMAC. We illustrate the evolution of the graph structure over the trajectory, as shown in the Figure \ref{f_7}. Initially, at $t=0$, no enemies are found within the allies' field of view, and the agents are unaware of each other's information. This condition result in a fully connected graph, indicating a state of complete communication. In the next timestep, if the agents' observations and states remain unchanged after their initial communication, the graph structure connections and communication volume decreased. At $t=3$, communication increas again when enemies entered the agents' fields of view. Similarly, in subsequent timesteps, minimal information changes lead to a synchronous reduction in communication. In summary, the visualization results indicate that the multi-key communication network learned by TGCNet possesses a certain level of interpretability.

\section{Conclusions and Future Work}

In this work, we propose TGCNet, a novel MARL algorithm that jointly learns communication policies and dynamic directed graph topologies to maximize team rewards. Our key contributions include formalizing a new paradigm for communicative, cooperative MARL that bridges training and execution through dynamic directed graphs, designing a Transformer-based multi-key communication mechanism for selective information transfer, leveraging graph coarsening networks to aggregate local observations, and demonstrating state-of-the-art performance on various cooperative MARL benchmarks. To the best of our knowledge, TGCNet is the first approach to avoid the input of global states through graph coarsening networks and to apply dynamic directed graphs to MARL. In the future, we plan to evaluate TGCNet on more complex real-world multi-agent domains, investigate emergent communication protocols and coordination behavior, and incorporate model-based planning and exploration techniques.

\appendix
\section*{Acknowledgments}
This work was supported in part by the National Natural Science Foundation of China under grants 62103302 and 62088101, by Shanghai Rising-Star Program under grant 24QA2709400, by the Shanghai Chenguang Program under grant 22CGA19, by the Shanghai Municipal Science and Technology Major Project under grant 2021SHZDZX0100, and by Fundamental Research Funds for the Central Universities under grant 22120240276. 

\bibliography{aaai25}

\end{document}